\documentclass[twocolumn,showpacs,aps,groupedaddress,prd,eqsecnum]{revtex4}

\usepackage{amsfonts,latexsym,eucal}
\usepackage[dvips]{graphicx}

\newcommand{\dalm}{\kern1pt\vbox{\hrule height 0.9pt\hbox{\vrule width 0.9pt
\hskip 2.5pt\vbox{\vskip 5.5pt}\hskip 3pt\vrule width 0.3pt}\hrule height 0.3pt}\kern1pt}

\usepackage{epsfig}

\begin{document}
% \draft command makes pacs numbers print 
% \draft

\thispagestyle{empty}

%%    -------
\title{Post-Newtonian parameters in the tensor-vector-scalar theory}
\author{Takashi Tamaki}
\email{tamaki@gravity.phys.waseda.ac.jp}
\address{Department of Physics, Waseda University, Okubo 3-4-1, Tokyo 169-8555, Japan}
%%    -------

%\date{\today}

\begin{abstract}
We investigate post-Newtonian parameters in the tensor-vector-scalar (TeVeS) theory in a general 
setting while previous researches have been restricted to spherically symmetric cases. 
Based on the assumption that both the physical and Einstein metrics have Minkowski metric 
at the zeroth order, we show $\gamma =1$ as in the previous researches. 
We find two remarkable things for other parameters. 
The first is the value $\beta =1$ while it has been reported 
that $\beta\neq 1$ for the case when the vector field is not purely timelike. 
This discrepancy occurs from the above assumption which is natural as a starting point. 
The second is the result that the Newtonian potential must be static to be consistent 
with the vector equation. As a result, we cannot determine $\alpha_{1}$ and $\alpha_{2}$. 
We consider that it is related to 
the instability against linear perturbation and occurrence of caustic 
singularities for various initial perturbations which have been reported recently.
\end{abstract}

\maketitle

%%%%%%%%%%%%%%%%%%%%%%
%%%%%%%%%%%%%%%%%%%%%%
\section{Introduction}
%%%%%%%%%%%%%%%%%%%%%%
%%%%%%%%%%%%%%%%%%%%%%

It is frequently argued that gravitational theories are an alternative 
to dark energy and dark matter. The tensor-vector-scalar (TeVeS) theory, 
which has been proposed by Bekenstein~\cite{Bekenstein}, is one of such theories and 
has attracted much attention because of its elegant way to explain 
galaxy rotation curves and the Tully-Fisher law. 
Thus, it is important to investigate its observational consequences for 
TeVeS to be a viable theory. 

One of the motivations of TeVeS is to explain the gravitational lensing 
without dark matter. 
As for strong gravitational lensing, possible explanation 
using only by TeVeS has been reported in \cite{Chen}. While 
the difficulty has been reported in explaining gravitational lensing 
of colliding galaxy clusters 1E0657-56 in modified gravities~\cite{Clowe}. 
Evolution of the universe and cosmological perturbation have also been studied 
in \cite{Chiu,Skordis,Skordis2}. 
These results tell us that TeVeS can explain the observed spectrum of the 
spatial distribution of galaxies and of the cosmic microwave radiation without 
dark matter if we include the contributions by neutrinos and the cosmological 
constant. The scalar field as an alternative to dark energy has been pointed out 
in \cite{Hao}, if the free function in TeVeS is chosen appropriately. 

However, it is not still evident that TeVeS can satisfy solar experiments. 
Parametrized post-Newtonian (PPN) parameters has been investigated in 
\cite{Bekenstein,Giannios} under the assumption of spherically symmetric 
space-time. For this reason, although their results show that $\gamma =1$, it does not 
guarantee the same value in a general setting. As for $\beta$, it has been 
reported that $\beta =1$ if the vector field is purely timelike and 
$\beta\neq 1$ if it is not. It is important to notice that the Einstein metric 
(We explain its meaning and the difference from the physical metric in Sec.II.) 
is different from the Minkowski metric at the zeroth order in \cite{Giannios}. 
In our opinion, it is the ambiguity peculiar to the bimetric theory since 
the requirement we can safely use is that the physical metric is Minkowski 
metric at the zeroth order. However, since the assumption that the Einstein metric 
is also Minkowski metric at the zeroth order is simple and hopeful from the aesthetic view, 
it is important to investigate its consequences. It is also important to investigate $\alpha_{1}$ 
and $\alpha_{2}$ which characterize the preferred frame effects. 
For these reasons we determine PPN parameters based on the general treatment 
in~\cite{Will}. We also follow \cite{EA-PPN}, which investigate PPN parameters in Einstein-Aether 
theory~\cite{start}, for our calculation and presentation. 
We obtain $\gamma =\beta =1$ as similar to the case 
where the radial component of the vector field is neglected. 
We cannot determine $\alpha_{1}$ and $\alpha_{2}$ because of the pathological 
situation that the Newtonian potential must be static. 

This paper is organized as follows. In Sec.~II, we explain TeVeS and give basic equations 
for later use. In Sec.~III, we summarize the set up for investigating PPN parameters. 
In Sec.~IV, we show the results and explain the difference from the previous 
result~\cite{Giannios}. In Sec.~V, consequences and future subjects are discussed. 
We use units in which $c=1$ 
and follow the sign conventions of Misner, Thorne, and
Wheeler~\cite{Misner:1974qy}, e.g., $(-,+,+,+)$ for metrics. 
We assume that Greek indices describe space-time as 
$\alpha, \beta =0,1,2,3$ and Roman indices describe space as $i,j =1,2,3$. 

%%%%%%%%%%%%%%%%%%%%%%%%%%%%%%%%%%%%%%%
%%%%%%%%%%%%%%%%%%%%%%%%%%%%%%%%%%%%%%%
\section{TeVeS and basic equations}
%%%%%%%%%%%%%%%%%%%%%%%%%%%%%%%%%%%%%%%
%%%%%%%%%%%%%%%%%%%%%%%%%%%%%%%%%%%%%%%
The action of TeVeS consists of four parts. We use the notation in 
\cite{Eva}. One is the Einstein-Hilbert 
action: 
%%%%%%%%%%%%%%%%%%%%%%%%%%
\begin{eqnarray}
\label{HEaction}
S_g =\frac{1}{16\pi G}\int
R\, \sqrt{-g}\,d^4x .
\end{eqnarray}
%%%%%%%%%%%%%%%%%%%%%%%%%%%%%
We express the metric of this action by $g_{\mu\nu}$ (the Einstein metric). 
The second is the vector field's action ($g_{\rm V}$ is a dimensionless positive 
coupling constant):
%%%%%%%%%%%%%%%%%%%%%%%%
\begin{eqnarray}
S_v&=&-\frac{g_{\rm V}}{32\pi
G}\int  \Big[\left(g^{\alpha\beta}g^{\mu\nu}u_{[\alpha,\mu]}u_{[\beta,\nu]}\right)
\nonumber \\
&&-\frac{2\lambda}{g_{\rm V}}\left(g^{\mu\nu}u_\mu
u_\nu+1\right)\Big]\,\sqrt{-g}\,d^4x,
\label{vectoraction}
\end{eqnarray}
%%%%%%%%%%%%%%%%%%%%%%%%
which includes a constraint that guarantees the vector field to be unit
timelike and $\lambda$ is the corresponding Lagrange multiplier 
and $u_{[\sigma,\,\mu]}:=u_{\sigma,\,\mu}-u_{\mu,\,\sigma}$.   
The third is the scalar's action ($k$ is a
dimensionless positive parameter while $\ell$ is a constant with the
dimensions of length, and  ${\cal F}$ a dimensionless free
function):
%%%%%%%%%%%%%%%%%%%%%%%%
\begin{eqnarray}
\label{scalaraction}S_s=-\frac{1}{2 k^2 \ell^2 G}\int
\mathcal{F}\left(k
\ell^2h^{\alpha\beta}\phi_{,\,\alpha}\phi_{,\,\beta}\right)\,\sqrt{-g}\,d^4x,
\end{eqnarray}
%%%%%%%%%%%%%%%%%%%%%%%%
where $h^{\alpha\beta}:= g^{\alpha\beta}-u^\alpha u^\beta$
with $u^\alpha := g^{\alpha\beta}u_\beta$. We assume that the 
vector and the scalar fields propagate along the Einstein metric. 

One of the points which characterizes TeVeS is the assumption that 
usual matter such as electromagnetic fields propagate 
along the physical metric $\tilde g_{\mu\nu}$ which is defined as 
%%%%%%%%%%%%%%%%%%
\begin{eqnarray}
\label{grelation}
\tilde{g}_{\alpha\beta}:=e^{-2\phi}g_{\alpha\beta}-2u_\alpha u_\beta
\sinh(2\phi). 
\end{eqnarray}
%%%%%%%%%%%%%%%%%%
This point is important to fit the gravitational lensing by 
clusters and galaxies with observations without dark matter. 
Accordingly, the matter action for field variables collectively denoted $f$ is:
%%%%%%%%%%%%%%%%%%%%%%%%%%
\begin{eqnarray}
S_m=\int \mathcal{L}\left(\tilde{g}_{\mu\nu},
f^\alpha,f^\alpha_{|\mu},\cdot\cdot\cdot\right)\, \sqrt{-\tilde{g}}\,d^4x\ ,
\label{matteraction}
\end{eqnarray}
%%%%%%%%%%%%%%%%%%%%%%%%%%%%%
where the covariant derivative denoted by $|$ is taken with respect to $\tilde{g}_{\mu\nu}$. 

Variation of the total action with respect to $g^{\alpha\beta}$
yields the TeVeS Einstein equations for $g_{\alpha\beta}$;
%%%%%%%%%%%%%%%%%%%%%%%%%%%
\begin{eqnarray}
\label{metric_eq} G_{\alpha\beta}&=&8\pi G\left(
\tilde{T}_{\alpha\beta}+\left(1-e^{-4\phi}\right)u^\mu
\tilde{T}_{\mu(\alpha}u_{\beta)}+\tau_{\alpha\beta}\right)
\nonumber  \\
&&+\theta_{\alpha\beta}\ .
\end{eqnarray} 
%%%%%%%%%%%%%%%%%%%%%%%%%%%
Here, $\tilde{T}_{\mu(\alpha}u_{\beta)}:=\tilde{T}_{\mu\alpha}u_{\beta}
+\tilde{T}_{\mu\beta}u_{\alpha}$. 
The sources here are the usual matter energy-momentum tensor
$\tilde T_{\alpha\beta}$, the variational derivative of $S_m$ with
respect to $\tilde g^{\alpha\beta}$, as well as  the energy-momentum
tensors for the scalar and vector fields:
%%%%%%%%%%%%%%%%%%%%%%%%%%%
\begin{eqnarray}
\tau_{\alpha\beta}&:=&
\frac{\mu}{kG}\left(\phi_{,\, \alpha}\phi_{, \,\beta}-u^\mu\phi_{,
\mu}u_{(\alpha}\phi_{, \,\beta)}\right)-\frac{\mathcal{F}
 g_{\alpha\beta}}{2k^2 \ell^2 G}\,,
\label{tau} \\
\nonumber
\theta_{\alpha\beta}&:=& g_{\rm V}\left(g^{\mu\nu}u_{[\mu,\,\alpha]}u_{[\nu,\,\beta]}-
\frac{1}{4}g^{\sigma\tau}g^{\mu\nu}u_{[\sigma,\,\mu]}u_{[\tau,\,\nu]}g_{\alpha\beta}\right)
\\
&-&\lambda u_\alpha u_\beta , \label{theta}
\end{eqnarray}
%%%%%%%%%%%%%%%%%%%%%%%%%%%
where
%%%%%%%%%%%%%%%%%%%%%
\begin{eqnarray}
\mu(x):=\frac{d\mathcal{F}(x)}{dx}\ .
\end{eqnarray}
%%%%%%%%%%%%%%%%%%%%%

The equations of
motion for the vector and scalar fields are, respectively,
%%%%%%%%%%%%%%%%%%%%%
\begin{eqnarray}
&&\left[\mu \Big(kl^2 h^{\gamma\delta} \phi_{,\,\gamma}\phi_{,\,\delta}\Big)
h^{\alpha\beta}\phi_{,\,\alpha}\right]_{;\,\beta}
\nonumber  \\
&=&kG\left[g^{\alpha\beta}+\left(1+e^{-4\phi}\right)u^\alpha
u^\beta\right] \tilde{T}_{\alpha\beta}\,,
\label{scalar_eq}  \\
&&g_{\rm V}u^{[\alpha;\beta]}\;_{;\beta}+\lambda u^\alpha+\frac{8\pi}{k}\mu
u^\beta\phi_{,\,\beta}g^{\alpha\gamma}\phi_{,\,\gamma}\nonumber
\\&=&8\pi
G\left(1-e^{-4\phi}\right)g^{\alpha\mu}u^\beta\tilde{T}_{\mu\beta}\ .
\label{vector_eq}
\end{eqnarray}
%%%%%%%%%%%%%%%%%%%%%%
There is the normalization condition on the vector field:
%%%%%%%%%%%%%%%%%%%%%%%
\begin{eqnarray}
\label{normalization}
u^\alpha u_\alpha=g_{\alpha\beta}\,u^\alpha
u^\beta=-1.  \label{constraint}
\end{eqnarray}
%%%%%%%%%%%%%%%%%%%%%%%
The Lagrange multiplier $\lambda$ can be calculated from the vector
equation.
%%%%%%%%%%%%%%%%%%%%%%%
\begin{eqnarray}
&&\lambda =g_{\rm V}u_{\alpha}u^{[\alpha;\beta]}\;_{;\beta}+\frac{8\pi}{k}
u^\beta\phi_{,\,\beta}u^{\gamma}\phi_{,\,\gamma}
\nonumber \\
&&-8\pi G\left(1-e^{-4\phi}\right)u^{\mu}u^\beta\tilde{T}_{\mu\beta}\ .
\label{lambda_eq}
\end{eqnarray}
%%%%%%%%%%%%%%%%%%%%%%%
By adjusting $\mathcal{F}$, we can explain galaxy rotation curves 
while preserving the Newtonian dynamics near the solar system. 
Below, we concentrate on the case which is responsible for 
the Newtonian dynamics, i.e., 
%%%%%%%%%%%%%%%%%%%%%%%
\begin{eqnarray}
\label{limit}
\mu =1\ .
\end{eqnarray}
%%%%%%%%%%%%%%%%%%%%%%%

%%%%%%%%%%%%%%%%%%%%%%%%%%%%%%%%%%%%%%%%%%%%%%%%%%%%%%%%%%%%%%%%%%%%%%%%
\section{set up}
%%%%%%%%%%%%%%%%%%%%%%%%%%%%%%%%%%%%%%%%%%%%%%%%%%%%%%%%%%%%%%%%%%%%%%%%
Here, we follow \cite{Will,EA-PPN} for our presentation. 

We assume the matter energy-momentum tensor as 
%%%%%%%%%%%%%%%%%%%%%%%
\begin{eqnarray}
\label{matter}
\tilde{T}^{\alpha\beta}=(\rho +\rho \Pi +p)v^{\alpha}v^{\beta}
+p\tilde{g}^{\alpha\beta}\ ,
\end{eqnarray}
%%%%%%%%%%%%%%%%%%%%%%%
where $v^{\alpha}$ is the four-velocity, $\rho$ the rest-mass energy density, 
$\Pi$ the internal energy density, and $p$ the isotropic pressure of the fluid. 

We assume a nearly globally Lorentzian coordinate system and both the physical metric 
and the Einstein metric are $\eta_{\alpha\beta}$, which is the basic difference from 
\cite{Giannios}, and the vector field is purely 
timelike at the zeroth order. 

As it is usual in the PPN formalism, we consider the slow motion and weak gravity 
limit which are characterized by 
%%%%%%%%%%%%%%%%%%%%%%%
\begin{eqnarray}
\frac{\partial}{\partial t}\sim O(0.5),\ \ 
\rho\sim\Pi\sim \frac{p}{\rho}\sim (v^{i})^{2}\sim O(1)\ . \label{basic-PPN}
\end{eqnarray}
%%%%%%%%%%%%%%%%%%%%%%%
We also assume that the components of the physical metric perturbations 
$\tilde{h}_{\alpha\beta}:=\tilde{g}_{\alpha\beta}-\eta_{\alpha\beta}$ will be 
%%%%%%%%%%%%%%%%%%%%%%%
\begin{eqnarray}
\tilde{h}_{00}\sim O(1),\ \ \tilde{h}_{ij}\sim O(1),\ \ 
\tilde{h}_{0i}\sim O(1.5). \label{expandh}
\end{eqnarray}
%%%%%%%%%%%%%%%%%%%%%%%
Notice that O(0.5) terms appear for the consistency with 
$\frac{\partial}{\partial t}\sim v^{i}\sim O(0.5)$. 
Notice also that these should be distinguished from 
the  Einstein metric perturbations $h_{\alpha\beta}$. 
Since we assumed that $g_{\alpha\beta}=\eta_{\alpha\beta}$ 
at the zeroth order, 
we can define $h_{\alpha\beta}:=g_{\alpha\beta}-\eta_{\alpha\beta}$. 
This means $h_{\alpha\beta}$ is same order as $\tilde{h}_{\alpha\beta}$ 
from the consistency with (\ref{grelation}), i.e.,
%%%%%%%%%%%%%%%%%%%%%%%
\begin{eqnarray}
h_{00}\sim O(1),\ \ h_{ij}\sim O(1),\ \ h_{0i}\sim O(1.5). \label{expandh2}
\end{eqnarray}
%%%%%%%%%%%%%%%%%%%%%%%
This makes important difference from \cite{Giannios}, since 
we have the vector perturbations $\delta u^{a}$ as 
%%%%%%%%%%%%%%%%%%%%%%%
\begin{eqnarray}
\delta u^{0}\sim O(1),\ \ \delta u^{i}\sim O(1.5)\ ,
\label{expandu}
\end{eqnarray}
%%%%%%%%%%%%%%%%%%%%%%%
to be consistent with (\ref{expandh2}). (In \cite{Giannios}, $\delta u^{i}\sim O(1)$ 
is assumed. One of the most efficient ways to see why (\ref{expandu}) appears is to 
see (\ref{ui-basic}). Lower order of $\delta u^{\alpha}$ cannot be permitted 
to be consistent with the Lorentz transformation and (\ref{expandh2}). 
I.e., since $h_{0i}\sim O(1.5)$, $u^i$ and $u_i$ are also $O(1.5)$. 
There is a possibility that $u^i$ and $u_i$ are lower order and 
$(u^i -u_i )\sim O(1.5)$ in some frame. 
However, cancellation in every Lorentz frame is impossible.) 
As for the scalar field, we have 
%%%%%%%%%%%%%%%%%%%%%%%
\begin{eqnarray}
\phi=\phi_{0}+O(1)\ .\label{expandphi}
\end{eqnarray}
%%%%%%%%%%%%%%%%%%%%%%%
Notice that we can set 
%%%%%%%%%%%%%%%%%%%%%%%
\begin{eqnarray}
\phi_{0}=0\ ,\label{phi0}
\end{eqnarray}
%%%%%%%%%%%%%%%%%%%%%%%
which is possible by rescaling the coordinates by factors $e^{\phi_{0}}$ 
using the relations (\ref{expandh}) and (\ref{expandu}). 
This means the consistency with the assumption that both the physical metric and the Einstein metric are 
$\eta_{\alpha\beta}$ at the zeroth order and characterize the difference from 
the $u^{r}\neq 0$ case in \cite{Giannios}. 

The physical metric components are expanded using particular potential 
functions which define the PPN parameters as
%%%%%%%%%%%%%%%%%%%%%%%
\begin{eqnarray}
&&\tilde{g}_{00}=-1+2U-2\beta U^{2}-2\xi \Phi_{\rm W}-(\zeta_{1}-2\xi ){\cal A} 
\nonumber  \\
&&+(2\gamma +2+\alpha_{3}+\zeta_{1}-2\xi )\Phi_{1}
\nonumber \\
&&+2(3\gamma -2\beta +1+\zeta_{2}+\xi )\Phi_{2}
\nonumber  \\
&&+2(1+\zeta_{3})\Phi_{3}
+2(3\gamma +3\zeta_{4}-2\xi )\Phi_{4}\ , 
\label{g00}  \\
&&\tilde{g}_{0j}=-\frac{1}{2}(4\gamma +3+\alpha_{1}-\alpha_{2}+\zeta_{1}-2\xi )V_{j}
\nonumber  \\
&&-\frac{1}{2}(1+\alpha_{2}-\zeta_{1}+2\xi )W_{j}\ , 
\label{g0i} \\
&&\tilde{g}_{jk}=(1+2\gamma U)\delta_{jk}\ .  \label{gjk}
\end{eqnarray}
%%%%%%%%%%%%%%%%%%%%%%%
The potentials are all of the form 
%%%%%%%%%%%%%%%%%%%%%%%
\begin{eqnarray}
F(x)=G_{\rm N}\int d^{3}y\frac{\rho (y)f}{|x-y|}\ ,
\end{eqnarray}
%%%%%%%%%%%%%%%%%%%%%%%
where $G_{\rm N}$ is the present value of Newton's constant, which we 
determine below in terms of $G$, $k$ and $g_{\rm V}$. $f$ characterizes 
each potentials. The correspondences $F:f$ are  
%%%%%%%%%%%%%%%%%%%%%%%
\begin{eqnarray}
&&U:1,\ \ \Phi_{1}:v_{i}v_{i},\ \ \Phi_{2}:U,\ \ \Phi_{3}:\Pi,\ \ \Phi_{4}:\frac{p}{\rho}, 
\nonumber \\
&&\Phi_{\rm W}:\int d^{3}z \rho (z)\frac{(x-y)_{j}}{|x-y|^{2}}
\left( \frac{(y-z)_{j}}{|y-z|} \right), \ \ V_{i}:v^{i},
\nonumber  \\
&&{\cal A}:\frac{[v_{i}(x-y)_{i}]^{2}}{|x-y|^{2}},\ \ 
W_{i}:\frac{v_{j}(x-y)_{j}(x-y)^{i}}{|x-y|^{2}}\ .
\end{eqnarray}
%%%%%%%%%%%%%%%%%%%%%%%
The relations useful below are 
%%%%%%%%%%%%%%%%%%%%%%%
\begin{eqnarray}
F_{,ii}=-4\pi G_{\rm N}\rho f\ , \label{func}
\end{eqnarray}
%%%%%%%%%%%%%%%%%%%%%%%
for $U$, $\Phi_{1,2,3,4}$, and $V_{i}$. 

We also define the superpotential $\chi$: 
%%%%%%%%%%%%%%%%%%%%%%%
\begin{eqnarray}
\chi =-G_{\rm N}\int d^{3}y\rho |x-y|\ ,
\end{eqnarray}
%%%%%%%%%%%%%%%%%%%%%%%
which satisfies 
%%%%%%%%%%%%%%%%%%%%%%%
\begin{eqnarray}
\chi_{,ii} =-2U\ .\label{super}
\end{eqnarray}
%%%%%%%%%%%%%%%%%%%%%%%
We also use the relation 
%%%%%%%%%%%%%%%%%%%%%%%
\begin{eqnarray}
\chi_{,0i} =V_{i}-W_{i}\ ,\label{super2}
\end{eqnarray}
%%%%%%%%%%%%%%%%%%%%%%%
which originally follows from the continuity equation for the fluid. 

There is a possibility that $\tilde{g}_{00}$ may depend on $\chi_{,00}$, 
and $\tilde{g}_{ij}$ may depend on $\chi_{,ij}$. However, it is known that 
these terms can be eliminated using gauge freedoms of the coordinate. 
To obtain this 
property, we shall impose the following gauge conditions: 
%%%%%%%%%%%%%%%%%%%%%%%
\begin{eqnarray}
h_{ij,j}&=&\frac{1}{2}(h_{jj,i}-h_{00,i})\ ,\label{gauge1}  \\
h_{0i,i}&=&AU_{,0}+Bu_{i,i}\ ,\label{gauge2}
\end{eqnarray}
%%%%%%%%%%%%%%%%%%%%%%%
where $A$ and $B$ are functions of $g_{\rm V}$ and $k$ which will be determined 
below. 

We also rewrite (\ref{metric_eq}) to the form
%%%%%%%%%%%%%%%%%%%%%%%%%%%
\begin{eqnarray}
\label{metric_eq2} R_{\alpha\beta}&=&\left[ 8\pi G\left(
\tilde{T}_{\mu\nu}+\left(1-e^{-4\phi}\right)u^\gamma
\tilde{T}_{\gamma (\mu}u_{\nu)}+\tau_{\mu\nu}\right)\right.
\nonumber  \\
&&+\left.\theta_{\mu\nu} \right](\delta_{\alpha}^{\mu}\delta_{\beta}^{\nu}
-\frac{1}{2}g_{\alpha\beta}g^{\mu\nu})\ .
\end{eqnarray} 
%%%%%%%%%%%%%%%%%%%%%%%%%%%

We determine PPN parameters along the following recipe:

(i) Solve the constraint (\ref{constraint}) for $u^{0}$ to O(1). 

(ii) Relate $h_{\mu\nu}$ with $\tilde{h}_{\mu\nu}$ using (\ref{grelation}) to O(1.5). 

(iii) Solve eq.(\ref{scalar_eq}) for $\phi$ to O(1). 

(iv) Calculate eq.(\ref{lambda_eq}) for $\lambda$ to O(1). 

(v) Solve $00$-component of eq.(\ref{metric_eq2}) for 
$\tilde{g}_{00}$ to O(1). 

(vi) Solve $ij$-components of eq.(\ref{metric_eq2}) for 
$\tilde{g}_{ij}$ to O(1). 

(vii) Solve $i$-components of eq.(\ref{vector_eq}) for $u^{i}$ to O(1.5). 

(viii) Solve $0i$-components of eq.(\ref{metric_eq2}) for 
$\tilde{g}_{0i}$ to O(1.5). 

(ix) Solve eq.(\ref{scalar_eq}) for $\phi$ to O(2). 

(x) Relate $h_{\mu\nu}$ with $\tilde{h}_{\mu\nu}$ using (\ref{grelation}) to O(2). 

(xi) Solve $00$-component of eq.(\ref{metric_eq2}) for 
$\tilde{g}_{00}$ to O(2). 

Below, we use the notation as, e.g., 
%%%%%%%%%%%%%%%%%%%%%%%
\begin{eqnarray}
\tilde{h}_{00}=\tilde{h}_{00}^{(1)}+\tilde{h}_{00}^{(2)}\ , 
\phi =\phi^{(1)}+\phi^{(2)}\ ,   
\end{eqnarray}
%%%%%%%%%%%%%%%%%%%%%%%
since we consider quantities to O(2), 
where the quantities of the superscript $(i)$ mean the quantities of order $i$. 

%%%%%%%%%%%%%%%%%%%%%%%%%%%%%%%%%%%%%%%
%%%%%%%%%%%%%%%%%%%%%%%%%%%%%%%%%%%%%%%
\section{calculation of PPN parameters}
%%%%%%%%%%%%%%%%%%%%%%%%%%%%%%%%%%%%%%%
%%%%%%%%%%%%%%%%%%%%%%%%%%%%%%%%%%%%%%%
%%%%%%%%%%%%%%%%%%%%%%%%%%%%%%%%%%%
\subsection{Solving O(1)}
%%%%%%%%%%%%%%%%%%%%%%%%%%%%%%%%%%%
{\bf (i):} From (\ref{normalization}), we obtain 
%%%%%%%%%%%%%%%%%%%%%%%
\begin{eqnarray}
u^{0}=1+\frac{h_{00}^{(1)}}{2}\ ,  \label{u0}
\end{eqnarray}
%%%%%%%%%%%%%%%%%%%%%%%
to O(1) where (\ref{expandh2}) and (\ref{expandu}) are used. From this, we have 
%%%%%%%%%%%%%%%%%%%%%%%
\begin{eqnarray}
u_{0}&=&g_{\alpha 0}u^{\alpha}=-1+\frac{h_{00}^{(1)}}{2}\ ,  \nonumber  \\
u_{i}&=&g_{\alpha i}u^{\alpha}=h_{0i}+u^{i}\ .\label{ui-basic}
\end{eqnarray}
%%%%%%%%%%%%%%%%%%%%%%%
From (\ref{normalization}), we also obtain 
%%%%%%%%%%%%%%%%%%%%%%%
\begin{eqnarray}
u^{0;\alpha}=0\ , \label{u-derivative}
\end{eqnarray}
%%%%%%%%%%%%%%%%%%%%%%%
to O(2). 

{\bf (ii):} From (\ref{grelation}), we obtain
%%%%%%%%%%%%%%%%%%%%%%%
\begin{eqnarray}
h_{00}^{(1)}=2\phi^{(1)}+\tilde{h}_{00}^{(1)}, \ \ 
h_{0i}=\tilde{h}_{0i},
\label{1-grelation}
\end{eqnarray}
%%%%%%%%%%%%%%%%%%%%%%%
using (\ref{grelation}), (\ref{expandphi}), (\ref{phi0}) and (\ref{u0}). 
Below, we also use 
%%%%%%%%%%%%%%%%%%%%%%%
\begin{eqnarray}
h_{00}^{(1)}=-h^{00(1)},\ \ 
h_{ij}^{(1)}=-h^{ij(1)}.\label{inverse}
\end{eqnarray}
%%%%%%%%%%%%%%%%%%%%%%%

{\bf (iii):} If we notice that $(\sqrt{-g})_{,\alpha}$ is the quantity of O(1) or 
higher, we can evaluate the l.h.s. of (\ref{scalar_eq}) to O(1) as 
%%%%%%%%%%%%%%%%%%%%%%%
\begin{eqnarray}
&&\frac{1}{\sqrt{-g}}[\sqrt{-g}(g^{\alpha\beta}-u^{\alpha}u^{\beta})
\phi_{,\alpha }]_{,\beta}
\nonumber  \\
&&=[(g^{\alpha\beta}-u^{\alpha}u^{\beta})
\phi_{,\alpha }]_{,\beta}=\phi_{,ii}^{(1)}\ ,
\label{lhs-scalar_eq}
\end{eqnarray}
%%%%%%%%%%%%%%%%%%%%%%%
where (\ref{basic-PPN}) are used. Then, we obtain 
%%%%%%%%%%%%%%%%%%%%%%%
\begin{eqnarray}
\phi_{,ii}^{(1)}=kG\rho\ .
\label{1-scalar_eq}
\end{eqnarray}
%%%%%%%%%%%%%%%%%%%%%%%
If we compare this with $U_{,ii}=-4\pi G_{\rm N}\rho$ 
which is one of (\ref{func}), we obtain 
%%%%%%%%%%%%%%%%%%%%%%%
\begin{eqnarray}
\phi^{(1)}=-\frac{kG}{4\pi G_{\rm N}}U\ .
\label{phi-U}
\end{eqnarray}
%%%%%%%%%%%%%%%%%%%%%%%

{\bf (iv):} From eq.(\ref{lambda_eq}) to O(1), we obtain 
%%%%%%%%%%%%%%%%%%%%%%%
\begin{eqnarray}
\lambda =g_{\rm V}u_{0}(u^{0;i}_{\ \ ;i}-u^{i;0}_{\ \ ;i})\ .
\label{lambda1}
\end{eqnarray}
%%%%%%%%%%%%%%%%%%%%%%%
Thus, it is necessary to calculate $u^{\alpha ;\beta}_{\ \ \ ;\gamma}$. 
From (\ref{expandu}), we obtain 
%%%%%%%%%%%%%%%%%%%%%%%
\begin{eqnarray}
\hspace{-3mm}u^{\alpha ;\beta}=g^{\beta\mu}\left[ u^{\alpha}_{\ ,\mu}+\frac{u^{0}}{2}
g^{\alpha\theta}(h_{0\theta ,\mu}+h_{\theta\mu ,0}-h_{0\mu ,\theta}) \right],
\label{covariantu-general}
\end{eqnarray}
%%%%%%%%%%%%%%%%%%%%%%%
to O(2). From this general expression, we have 
%%%%%%%%%%%%%%%%%%%%%%%
\begin{eqnarray}
&&u^{i;0}=\frac{h_{00,i}}{2}\ ,  \label{u-1derivative1} \\ 
&&u^{i;k}=u_{i,k}+\frac{1}{2}(h_{ik,0}-h_{k0,i}-h_{i0,k})\ ,
\label{uik}
\end{eqnarray}
%%%%%%%%%%%%%%%%%%%%%%%
to O(1) and O(1.5), respectively. Here, we used (\ref{ui-basic}). 
From (\ref{u-derivative}), (\ref{u-1derivative1}), (\ref{uik}) 
and that $\Gamma^{\alpha}_{\beta\gamma}$ is O(1) or higher order quantity, 
we obtain 
%%%%%%%%%%%%%%%%%%%%%%%
\begin{eqnarray}
u^{\alpha ;\beta}_{\ \ \ ;\gamma}=u^{\alpha ;\beta}_{\ \ \ ,\gamma}\ , 
\label{usual-covariant}
\end{eqnarray}
%%%%%%%%%%%%%%%%%%%%%%%
to O(1.5). Using (\ref{ui-basic}), (\ref{u-derivative}), (\ref{u-1derivative1}) and 
(\ref{usual-covariant}) to (\ref{lambda1}), we obtain 
%%%%%%%%%%%%%%%%%%%%%%%
\begin{eqnarray}
\lambda =\frac{g_{\rm V}}{2}h_{00,ii}^{(1)}\ ,
\label{lambda2}
\end{eqnarray}
%%%%%%%%%%%%%%%%%%%%%%%
to O(1). 

{\bf (v):} To solve (\ref{metric_eq2}), we should evaluate each terms. 
As for l.h.s. of (\ref{metric_eq2}), we obtain
%%%%%%%%%%%%%%%%%%%%%%%
\begin{eqnarray}
&&R_{00}=\frac{1}{2}h_{ij}h_{00,ij}-\frac{1}{2}h_{00,ii}+
(h_{i0,i}-\frac{1}{2}h_{ii,0})_{,0}  \nonumber  \\
&&-\frac{1}{4}(h_{00,i})^{2}+\frac{1}{4}h_{00,j}(2h_{ij,i}-h_{ii,j})\ , 
\label{R00-order2}
\end{eqnarray}
%%%%%%%%%%%%%%%%%%%%%%%
to O(2). Thus, we have 
%%%%%%%%%%%%%%%%%%%%%%%
\begin{eqnarray}
R_{00}=-\frac{1}{2}h_{00,ii}^{(1)}\ ,
\label{R00}
\end{eqnarray}
%%%%%%%%%%%%%%%%%%%%%%%
to O(1). From (\ref{tau}), we obtain
%%%%%%%%%%%%%%%%%%%%%%%
\begin{eqnarray}
&&\tau_{00}=\frac{(\phi_{,k}^{(1)})^{2}}{2kG}\ ,\ \ \tau_{0i}=0\ ,  
\label{tau1}  \\
&&\tau_{ij}=\frac{1}{kG}\left[ \phi_{,i}^{(1)}\phi_{,j}^{(1)}-\frac{\delta_{ij}}{2}
(\phi_{,k}^{(1)})^{2}\right]  ,\label{tau2}
\end{eqnarray}
%%%%%%%%%%%%%%%%%%%%%%%
to O(2). We notice that $\tau_{\alpha\beta}$ disappears to O(1). 
From (\ref{matter}), we obtain 
%%%%%%%%%%%%%%%%%%%%%%%
\begin{eqnarray}
&&(1-e^{-4\phi})u^{\mu}\tilde{T}_{\mu (0}u_{0)}-
\frac{g_{00}}{2}(1-e^{-4\phi})u^{\mu}\tilde{T}_{\mu (\beta}u^{\beta )} \nonumber  \\
&&=-4\phi^{(1)}\rho\ ,
\label{2-sEM}
\end{eqnarray}
%%%%%%%%%%%%%%%%%%%%%%%
to O(2). Thus, this term also disappears to O(1). Then, 
if we notice that 
%%%%%%%%%%%%%%%%%%%%%%%
\begin{eqnarray}
&&\theta_{00}=-\lambda\ ,\ \ \ \theta_{0i}=\theta_{ij}=0\ , 
\label{1-theta}
\end{eqnarray}
%%%%%%%%%%%%%%%%%%%%%%%
to O(1), we obtain 
%%%%%%%%%%%%%%%%%%%%%%%
\begin{eqnarray}
\left( 1-\frac{g_{\rm V}}{2}\right)h_{00,ii}^{(1)}=-8\pi G\rho\ ,
\label{1-basic}
\end{eqnarray}
%%%%%%%%%%%%%%%%%%%%%%%
where (\ref{matter}), (\ref{lambda2}) and (\ref{R00}) are used. 
Using (\ref{1-grelation}) and (\ref{1-scalar_eq}), 
we have 
%%%%%%%%%%%%%%%%%%%%%%%
\begin{eqnarray}
\tilde{h}_{00,ii}^{(1)}=-8\pi G\rho\left[\left( 1-\frac{g_{\rm V}}{2}\right)^{-1}
+\frac{k}{4\pi}
\right]\ .\label{Newton1}
\end{eqnarray}
%%%%%%%%%%%%%%%%%%%%%%%
Since the physical metric is $\tilde{g}_{\mu\nu}$, the requirement that 
we recover Newton gravity is expressed as 
%%%%%%%%%%%%%%%%%%%%%%%
\begin{eqnarray}
\tilde{h}_{00}^{(1)}:=2U\ .\label{Newton2}
\end{eqnarray}
%%%%%%%%%%%%%%%%%%%%%%%
(\ref{Newton1}) and (\ref{Newton2}) tell us that 
%%%%%%%%%%%%%%%%%%%%%%%
\begin{eqnarray}
G_{\rm N}=G\left[\left( 1-\frac{g_{\rm V}}{2}\right)^{-1}
+\frac{k}{4\pi}\right]\ .\label{Newton}
\end{eqnarray}
%%%%%%%%%%%%%%%%%%%%%%%
This result is important since the current value of the gravitational 
constant could change its sign depending on $g_{\rm V}$ and $k$. 
This result is also consistent with \cite{Sagi2} except the normalization. 
We also have 
%%%%%%%%%%%%%%%%%%%%%%%
\begin{eqnarray}
h_{00}^{(1)}=2U\frac{G}{G_{\rm N}}\left(1-\frac{g_{\rm V}}{2}\right)^{-1}\ .  
\label{h00-U}
\end{eqnarray}
%%%%%%%%%%%%%%%%%%%%%%%

{\bf (vi):} If we use the gauge (\ref{gauge1}), 
we obtain $R_{ij}=-\frac{1}{2}h_{ij,kk}^{(1)}$. Then, we have 
%%%%%%%%%%%%%%%%%%%%%%%
\begin{eqnarray}
\tilde{h}_{ij,kk}^{(1)}=-8\pi G\rho\left[\left( 1-\frac{g_{\rm V}}{2}\right)^{-1}
+\frac{k}{4\pi}\right]\delta_{ij}\ ,
\end{eqnarray}
%%%%%%%%%%%%%%%%%%%%%%%
similar to the $00$-component. This means 
%%%%%%%%%%%%%%%%%%%%%%%
\begin{eqnarray}
\tilde{h}_{ij}^{(1)}=\tilde{h}_{00}^{(1)}\delta_{ij}\ ,\ \ 
h_{ij}^{(1)}=h_{00}^{(1)}\delta_{ij}\ .  \label{standard-PPN}
\end{eqnarray}
%%%%%%%%%%%%%%%%%%%%%%%
Using (\ref{Newton2}), we obtain 
%%%%%%%%%%%%%%%%%%%%%%%
\begin{eqnarray}
\gamma =1\ .
\end{eqnarray}
%%%%%%%%%%%%%%%%%%%%%%%
This means that $\gamma$ has same value as in the spherically 
symmetric case. Thus, we can conclude that TeVeS has not been 
excluded from the observation related to $\gamma$. 

%%%%%%%%%%%%%%%%%%%%%%%%%%%%%%%%%%%
\subsection{Solving O(1.5)}
%%%%%%%%%%%%%%%%%%%%%%%%%%%%%%%%%%%
{\bf (vii):} From $i$-component of eq.(\ref{vector_eq}), we obtain 
$u^{i;\beta}_{\ \ \ ;\beta}=u^{\beta ;i}_{\ \ \ ;\beta}$ to O(1.5) where we 
used $\lambda \sim O(1)$ as seen from (\ref{lambda2}). 
If we remember (\ref{u-derivative}) and (\ref{usual-covariant}), this means 
%%%%%%%%%%%%%%%%%%%%%%%
\begin{eqnarray}
u^{i;\beta}_{\ \ \ ,\beta}=u^{j ;i}_{\ \ \ ,j}\ .  
\label{metric1.5}
\end{eqnarray}
%%%%%%%%%%%%%%%%%%%%%%%
Then, we obtain 
%%%%%%%%%%%%%%%%%%%%%%%
\begin{eqnarray}
h_{00,i0}^{(1)}+2u_{i,jj}=2u_{j,ij}\ ,  \label{vector_1.5}
\end{eqnarray}
%%%%%%%%%%%%%%%%%%%%%%%
where we used (\ref{u-1derivative1}) and (\ref{uik}). By taking the 
spatial divergence of this equation, we obtain 
%%%%%%%%%%%%%%%%%%%%%%%
\begin{eqnarray}
h_{00,i0i}^{(1)}=0\ .  \label{static}
\end{eqnarray}
%%%%%%%%%%%%%%%%%%%%%%%
Thus, we conclude 
%%%%%%%%%%%%%%%%%%%%%%%
\begin{eqnarray}
h_{00,0}^{(1)}=0\ .  \label{static2}
\end{eqnarray}
%%%%%%%%%%%%%%%%%%%%%%%
This is very important since we obtain 
%%%%%%%%%%%%%%%%%%%%%%%
\begin{eqnarray}
U_{,0}=0\ ,  \label{static3}
\end{eqnarray}
%%%%%%%%%%%%%%%%%%%%%%%
from (\ref{h00-U}). This means that if we consider the perturbation 
around Minkowski space-time, the Newtonian potential must be static. 
This does not happen in the usual scalar-tensor theory~\cite{Will}. 
It is also important to notice that Einstein-Aether theory has a same 
property if we adopt a Maxwell type vector action~\cite{EA-PPN}. 
(\ref{static3}) also means that 
%%%%%%%%%%%%%%%%%%%%%%%
\begin{eqnarray}
\phi_{,0}^{(1)}=\chi_{,ii0}=0\ ,\ \   \label{static4}
\end{eqnarray}
%%%%%%%%%%%%%%%%%%%%%%%
from (\ref{super}) and (\ref{phi-U}). 

{\bf (viii):} By solving $0i$ components of eq.(\ref{metric_eq2}), 
we obtain 
%%%%%%%%%%%%%%%%%%%%%%%
\begin{eqnarray}
h_{0i,jj}-h_{0j,ij}=16\pi G\rho v_{i}\ ,  \label{metric_1.5}
\end{eqnarray}
%%%%%%%%%%%%%%%%%%%%%%% 
where we used (\ref{gauge1}) and (\ref{static2}). 
Using (\ref{gauge2}), (\ref{vector_1.5}) and (\ref{static3}), 
we have 
%%%%%%%%%%%%%%%%%%%%%%%
\begin{eqnarray}
h_{0j,ij}=Bu_{i,jj}\ .  \label{step1}
\end{eqnarray}
%%%%%%%%%%%%%%%%%%%%%%% 
By substituting this to (\ref{metric_1.5}) and using (\ref{vector_1.5}), we obtain 
%%%%%%%%%%%%%%%%%%%%%%%
\begin{eqnarray}
&&16\pi G\rho v_{i}=h_{0i,jj}-Bu_{i,jj}\ .  
\label{u-determined}
\end{eqnarray}
%%%%%%%%%%%%%%%%%%%%%%% 
Thus, $u_{i}$ is expressed as 
%%%%%%%%%%%%%%%%%%%%%%%
\begin{eqnarray}
&&Bu_{i}=h_{0i}+4\frac{G}{G_{\rm N}}V_{i}\ .  \label{u-determined2}
\end{eqnarray}
%%%%%%%%%%%%%%%%%%%%%%% 

%%%%%%%%%%%%%%%%%%%%%%%%%%%%%%%%%%%
\subsection{Solving O(2)}
%%%%%%%%%%%%%%%%%%%%%%%%%%%%%%%%%%%

{\bf (ix):} From (\ref{grelation}), we have 
%%%%%%%%%%%%%%%%%%%%%%%
\begin{eqnarray}
h_{00}^{(2)}=2\phi^{(2)}+\tilde{h}_{00}^{(2)}-2(\phi^{(1)})^{2}
-2\phi^{(1)}\tilde{h}_{00}^{(1)}\ .
\label{2-grelation}
\end{eqnarray}
%%%%%%%%%%%%%%%%%%%%%%%
This means that we need to solve (\ref{scalar_eq}) for $\phi^{(2)}$, 
which will be performed below, to determine $\beta$.

{\bf (x):} If we notice that
%%%%%%%%%%%%%%%%%%%%%%%
\begin{eqnarray}
&&\tilde{T}_{00}=\rho (1+\Pi +v_{i}v_{i}-2U)\ ,  \label{T00}  \\
&&\tilde{T}_{ij}=\rho v_{i}v_{j}+p\delta_{ij}\ , \label{Tij}
\end{eqnarray}
%%%%%%%%%%%%%%%%%%%%%%%
r.h.s. of (\ref{scalar_eq}) to O(2) is evaluated as
%%%%%%%%%%%%%%%%%%%%%%%
\begin{eqnarray}
\hspace{-8mm}kG\rho (1+\Pi +2v_{i}v_{i}-2U+3\frac{P}{\rho }+h_{00}^{(1)}-4\phi^{(1)})\ .
\label{rhs-scalar_eq2}
\end{eqnarray}
%%%%%%%%%%%%%%%%%%%%%%%
If we use (\ref{standard-PPN}), $\sqrt{-g}$ is evaluated as 
%%%%%%%%%%%%%%%%%%%%%%%
\begin{eqnarray}
\sqrt{-g}=1+h_{00}^{(1)}\ ,\label{sqrtg}
\end{eqnarray}
%%%%%%%%%%%%%%%%%%%%%%%
to O(1). Using (\ref{inverse}), (\ref{static4}) and (\ref{sqrtg}), 
l.h.s. of (\ref{scalar_eq}) to O(2) is calculated as 
%%%%%%%%%%%%%%%%%%%%%%%
\begin{eqnarray}
&&\frac{1}{\sqrt{-g}}[\sqrt{-g}(g^{\alpha\beta}-u^{\alpha}u^{\beta})
\phi_{,\alpha }]_{,\beta}
\nonumber  \\
&&=\frac{(\sqrt{-g})_{,i}}{\sqrt{-g}}[(g^{ij}-u^{i}u^{j})
\phi_{,j }]+(g^{ij}-u^{i}u^{j})_{,i}\phi_{,j }  
\nonumber  \\
&&+(g^{ij}-u^{i}u^{j})
\phi_{,ij }
\nonumber  \\
&&=\phi_{,ii}-h_{00}^{(1)}\phi_{,ii}^{(1)}\ .
\label{lhs-scalar_eq2}
\end{eqnarray}
%%%%%%%%%%%%%%%%%%%%%%%
Using (\ref{1-grelation}) and (\ref{1-scalar_eq}) to (\ref{rhs-scalar_eq2}) and 
(\ref{lhs-scalar_eq2}), we obtain 
%%%%%%%%%%%%%%%%%%%%%%%
\begin{eqnarray}
\phi_{,ii}^{(2)}=kG\rho (1+\Pi +2v_{i}v_{i}-2U+3\frac{P}{\rho }+2\tilde{h}_{00}^{(1)})\ .  
\label{2-scalar_eq}
\end{eqnarray}
%%%%%%%%%%%%%%%%%%%%%%%

{\bf (xi):} To evaluate r.h.s of eq.(\ref{metric_eq2}) to O(2), 
we need to evaluate each terms to O(2). Since we have (\ref{tau1}), 
(\ref{tau2}) and (\ref{2-sEM}) to O(2), the term remained to be evaluated 
is $\theta_{\mu\nu}$. By looking at (\ref{theta}), we notice 
that we need to evaluate $\lambda$ to O(2). Then, from eq.(\ref{lambda_eq}), 
we notice that we need to evaluate $u^{[\alpha;\beta ]}_{\ \ \ ;\beta}$ 
to O(2) which is one of the most tedious part in our calculation. 
Using (\ref{uik}) (quantity of O(1.5)) and (\ref{u-derivative}), 
we obtain 
%%%%%%%%%%%%%%%%%%%%%%%
\begin{eqnarray}
u^{\alpha;\beta}_{\ \ ;\gamma}=u^{\alpha ;\beta}_{\ \ ,\gamma}
+\Gamma^{\alpha}_{i\gamma}u^{i;\beta}+
\Gamma^{\beta}_{0\gamma}u^{\alpha ;0}\ .  \label{use-0}
\end{eqnarray}
%%%%%%%%%%%%%%%%%%%%%%%
From this expression, (\ref{uik}) (quantity of O(1.5)) and 
(\ref{u-derivative}), we obtain 
%%%%%%%%%%%%%%%%%%%%%%%
\begin{eqnarray}
u^{0;i}_{\ \ \ ;i}=u^{0;i}_{\ \ \ ,i}+\Gamma^{0}_{ij}u^{i;j}
+\Gamma^{i}_{0i}u^{0;0}=0\ ,  \label{use-1}
\end{eqnarray}
%%%%%%%%%%%%%%%%%%%%%%%
to O(2). Using (\ref{standard-PPN}), we have 
%%%%%%%%%%%%%%%%%%%%%%%
\begin{eqnarray}
u^{i;0}=-u_{i,0}+\frac{h_{00,i}}{2}+\frac{h_{00}^{(1)}h_{00,i}^{(1)}}{4}\ .  
\label{ui0}  
\end{eqnarray}
%%%%%%%%%%%%%%%%%%%%%%%
From (\ref{use-0}) and (\ref{ui0}), we obtain 
%%%%%%%%%%%%%%%%%%%%%%%
\begin{eqnarray}
&&u^{i;0}_{\ \ \ ;i}=\frac{1}{2}h_{00,ii}+\frac{3}{8}\left[(h_{00}^{(1)})^{2}
\right]_{,ii}
\nonumber  \\
&&-\frac{h_{00}^{(1)}h_{00,ii}^{(1)}}{2}-u_{i,0i}\ ,  \label{use-2}
\end{eqnarray}
%%%%%%%%%%%%%%%%%%%%%%%
where we used
%%%%%%%%%%%%%%%%%%%%%%%
\begin{eqnarray}
(h_{00,i}^{(1)})^{2}=\frac{1}{2}\left[(h_{00}^{(1)})^{2}
\right]_{,ii}-h_{00}^{(1)}h_{00,ii}^{(1)}\ .  \label{user-3}
\end{eqnarray}
%%%%%%%%%%%%%%%%%%%%%%%
Then, from (\ref{use-1}) and (\ref{use-2}), we can evaluate $\lambda$ to O(2) as 
%%%%%%%%%%%%%%%%%%%%%%%
\begin{eqnarray}
&&\lambda =g_{\rm V}u_{\alpha}u^{[\alpha ;\beta]}_{\ \ \ ;\beta}-32\pi G\phi^{(1)}\rho
\nonumber  \\
&&=-g_{\rm V}u_{0}u^{i;0}_{\ \ \ ;i}-32\pi G\phi^{(1)}\rho  
\nonumber  \\
&&=g_{\rm V}\left[ \frac{h_{00,ii}}{2}+\frac{3}{8}\left[(h_{00}^{(1)})^{2}
\right]_{,ii}\right.\nonumber  \\
&&\left. -u_{i,0i}-\frac{3}{4}h_{00}^{(1)}h_{00,ii}^{(1)} \right]
-32\pi G\phi^{(1)}\rho \ .   \label{use-3}
\end{eqnarray}
%%%%%%%%%%%%%%%%%%%%%%%
Using this result, we obtain the contribution of $\theta_{\mu\nu}$ for 
$00$-component of eq.(\ref{metric_eq2}) as 
%%%%%%%%%%%%%%%%%%%%%%%
\begin{eqnarray}
&&\theta_{00}-\frac{g_{00}}{2}\theta_{\alpha\beta}g^{\alpha\beta}=
\frac{1}{2}(\theta_{00}+\theta_{ii})  \nonumber  \\
&=&\frac{g_{\rm V}}{2}(u_{0,i})^{2}-\frac{\lambda}{2}(u_{0})^{2}  \nonumber  \\
&=&\frac{g_{\rm V}}{2}h_{00}^{(1)}h_{00,ii}^{(1)}-\frac{g_{\rm V}}{8}
\left[(h_{00}^{(1)})^{2}\right]_{,ii}+16\pi G\phi^{(1)}\rho   \nonumber  \\
&&+\frac{g_{\rm V}}{2}
(u_{i,0i}-\frac{1}{2}h_{00,ii})\ . \label{2-theta}
\end{eqnarray}
%%%%%%%%%%%%%%%%%%%%%%%
In a similar way, we obtain 
%%%%%%%%%%%%%%%%%%%%%%%
\begin{eqnarray}
&&\tilde{T}_{00}-\frac{g_{00}}{2}\tilde{T}_{\alpha\beta}g^{\alpha\beta}
=\frac{1}{2}(\tilde{T}_{00}+\tilde{T}_{ii})  \nonumber  \\
&&=\frac{\rho}{2}(1+\Pi +2v_{i}v_{i}-2U+3\frac{p}{\rho})\ ,
\label{2-EM}
\end{eqnarray}
%%%%%%%%%%%%%%%%%%%%%%%
and 
%%%%%%%%%%%%%%%%%%%%%%%
\begin{eqnarray}
&&\tau_{00}-\frac{g_{00}}{2}\tau_{\alpha\beta}g^{\alpha\beta}=0\ .
\label{2-scalar}
\end{eqnarray}
%%%%%%%%%%%%%%%%%%%%%%%
Thus, all quantities of r.h.s. of eq.(\ref{metric_eq2}) are evaluated. 
As for l.h.s., if we use (\ref{standard-PPN}), (\ref{static2}) and (\ref{user-3}) 
to (\ref{R00-order2}), we have 
%%%%%%%%%%%%%%%%%%%%%%%
\begin{eqnarray}
\hspace{-5mm}R_{00}=h_{00}^{(1)}h_{00,ii}^{(1)}-\frac{1}{2}h_{00,ii}+h_{i0,i0}
-\frac{1}{4}\left[(h_{00}^{(1)})^{2}
\right]_{,ii}. \label{2-eeq}
\end{eqnarray}
%%%%%%%%%%%%%%%%%%%%%%%
Therefore, by summarizing the results (\ref{2-sEM}), (\ref{2-theta}), (\ref{2-EM}), 
(\ref{2-scalar}) and (\ref{2-eeq}), we obtain 
%%%%%%%%%%%%%%%%%%%%%%%
\begin{eqnarray}
\hspace{-8mm}&&h_{00,ii}-2\left(1-\frac{g_{\rm V}}{2}\right)^{-1}(h_{i0,i0}
-\frac{g_{\rm V}}{2}u_{i,0i}) \nonumber  \\
\hspace{-8mm}&&+\frac{1}{2}\left[(h_{00}^{(1)})^{2}\right]_{,ii}-2h_{00}^{(1)}h_{00,ii}^{(1)}
-32\pi G\phi^{(1)}\rho \left(1-\frac{g_{\rm V}}{2}\right)^{-1} \nonumber  \\
\hspace{-8mm}&&=-8\pi G\rho \left(1-\frac{g_{\rm V}}{2}\right)^{-1}
(1+\Pi +2v_{i}v_{i}-2U+3\frac{p}{\rho }).
\label{2-basic}
\end{eqnarray}
%%%%%%%%%%%%%%%%%%%%%%%
If we use (\ref{1-basic}), (\ref{2-grelation}) and (\ref{2-scalar_eq}), we obtain 
%%%%%%%%%%%%%%%%%%%%%%%
\begin{eqnarray}
&&\tilde{h}_{00,ii}^{(2)}+4\phi_{,ii}^{(1)}\tilde{h}_{00}^{(1)}
-2\left[(\phi^{(1)})^{2}+\phi^{(1)}\tilde{h}_{00}^{(1)}\right]_{,ii}\nonumber  \\
&&+\frac{1}{2}\left[(h_{00}^{(1)})^{2}\right]_{,ii}-2h_{00}^{(1)}h_{00,ii}^{(1)}
-32\pi G\phi^{(1)}\rho \left(1-\frac{g_{\rm V}}{2}\right)^{-1} \nonumber  \\
&&-2\left(1-\frac{g_{\rm V}}{2}\right)^{-1}(h_{i0,i0}
-\frac{g_{\rm V}}{2}u_{i,0i})  \nonumber  \\
&&=-8\pi G_{\rm N}\rho (\Pi +2v_{i}v_{i}-2U+3\frac{p}{\rho })\ .
\label{2-basic2}
\end{eqnarray}
%%%%%%%%%%%%%%%%%%%%%%%
If we use (\ref{gauge2}), (\ref{phi-U}) and (\ref{h00-U}), 
we can rewrite this equation as 
%%%%%%%%%%%%%%%%%%%%%%%
\begin{eqnarray}
\hspace{-5mm}&&\left\{\tilde{h}_{00}^{(2)}-\frac{2kG}{\pi G_{\rm N}}\Phi_{2}-2\left(
\frac{kG}{4\pi G_{\rm N}}\right)^{2}U^{2}+\frac{kG}{\pi G_{\rm N}}U^{2}+\right. \nonumber  \\
\hspace{-5mm}&&\left.\left(\frac{G}{G_{\rm N}}\right)^{2}
\left(1-\frac{g_{\rm V}}{2}\right)^{-2}\left[2U^{2}
-8\Phi_{2}-\frac{2k}{\pi}\left(1-\frac{g_{\rm V}}{2}\right)\Phi_{2}\right]
\right\}_{,ii} \nonumber  \\
\hspace{-5mm}&&-2\left(1-\frac{g_{\rm V}}{2}\right)^{-1}\left(B-\frac{g_{\rm V}}{2}\right)u_{i,i0}
\nonumber  \\
\hspace{-5mm}&&=\left(2\Phi_{3}+4\Phi_{1}-4\Phi_{2}+6\Phi_{4}\right)_{,ii}\ .
\label{result1}
\end{eqnarray}
%%%%%%%%%%%%%%%%%%%%%%%
If we remember (\ref{u-determined2}), we notice that the coefficients of 
$u_{i,i0}$ should vanish for $\tilde{h}_{00}^{(2)}$ 
to be expressed without $h_{i0}$. Thus, we have 
%%%%%%%%%%%%%%%%%%%%%%%
\begin{eqnarray}
B=\frac{g_{\rm V}}{2}\ .\label{B}
\end{eqnarray}
%%%%%%%%%%%%%%%%%%%%%%%
Then, we obtain 
%%%%%%%%%%%%%%%%%%%%%%%
\begin{eqnarray}
&&\tilde{h}_{00}^{(2)}-\frac{2kG}{\pi G_{\rm N}}\Phi_{2}-2\left(
\frac{kG}{4\pi G_{\rm N}}\right)^{2}U^{2}+\frac{kG}{\pi G_{\rm N}}U^{2}+ \nonumber  \\
&&\left(\frac{G}{G_{\rm N}}\right)^{2}
\left(1-\frac{g_{\rm V}}{2}\right)^{-2}\left[2U^{2}
-8\Phi_{2}-\frac{2k}{\pi}\left(1-\frac{g_{\rm V}}{2}\right)\right] \nonumber  \\
&&=2\Phi_{3}+4\Phi_{1}-4\Phi_{2}+6\Phi_{4}\ .
\label{result}
\end{eqnarray}
%%%%%%%%%%%%%%%%%%%%%%%
As a result, we obtain 
%%%%%%%%%%%%%%%%%%%%%%%
\begin{eqnarray}
&&\ \ \xi =\zeta_{1}=\zeta_{3}=\zeta_{4}=\alpha_{3}=0\ ,\nonumber  \\
&&2\beta =\frac{kG}{\pi G_{\rm N}}-2\left(\frac{kG}{4\pi G_{\rm N}}\right)^{2}
\nonumber \\
&&+2\left(\frac{G}{G_{\rm N}}\right)^{2}
\left(1-\frac{g_{\rm V}}{2}\right)^{-2}\ .
\label{parameters}
\end{eqnarray}
%%%%%%%%%%%%%%%%%%%%%%%
Surprisingly, if we substitute (\ref{Newton}), we obtain 
%%%%%%%%%%%%%%%%%%%%%%%
\begin{eqnarray}
\beta =1.
\label{beta}
\end{eqnarray}
%%%%%%%%%%%%%%%%%%%%%%%
This is one of the major results in this paper. Thus, we cannot restrict 
TeVeS related to the experiment about $\beta$. 
From the coefficient of $\Phi_{2}$, we obtain 
%%%%%%%%%%%%%%%%%%%%%%%
\begin{eqnarray}
&&2(4-2\beta +\zeta_{2})=-4+\frac{2kG}{\pi G_{\rm N}}
+8\left(\frac{G}{G_{\rm N}}\right)^{2}
\left(1-\frac{g_{\rm V}}{2}\right)^{-2}
\nonumber  \\
&&+\frac{2k}{\pi}\left(\frac{G}{G_{\rm N}}\right)^{2}
\left(1-\frac{g_{\rm V}}{2}\right)^{-1}\ .
\label{parameters2}
\end{eqnarray}
%%%%%%%%%%%%%%%%%%%%%%%
By substituting (\ref{Newton}), we have 
%%%%%%%%%%%%%%%%%%%%%%%
\begin{eqnarray}
\zeta_{2}=0.
\label{parameters3}
\end{eqnarray}
%%%%%%%%%%%%%%%%%%%%%%%

Unfortunately, we cannot determine $\alpha_{1}$ and $\alpha_{2}$ since we cannot 
eliminate $u_{i}$ from (\ref{u-determined2}). 
Notice that although we apparently obtain 
%%%%%%%%%%%%%%%%%%%%%%%
\begin{eqnarray}
\tilde{h}_{0i}=-4\frac{G}{G_{\rm N}}V_{i},
\label{h0i}
\end{eqnarray}
%%%%%%%%%%%%%%%%%%%%%%%
for $g_{\rm V}=0$, where (\ref{1-grelation}) is used. We cannot use this equation to 
determine $\alpha_{1}$ and $\alpha_{2}$ since we use (\ref{metric1.5}), 
which is obtained from (\ref{vector_eq}) under the assumption $g_{\rm V}\neq 0$, 
to derive (\ref{h0i}).

%%%%%%%%%%%%%%%%%%%%%%%%%%%%%%%%%%%
%%%%%%%%%%%%%%%%%%%%%%%%%%%%%%%%%%%
\section{Conclusion and discussion}
%%%%%%%%%%%%%%%%%%%%%%%%%%%%%%%%%%%
%%%%%%%%%%%%%%%%%%%%%%%%%%%%%%%%%%%
We have investigated PPN parameters of TeVeS from the general procedure which 
has not been done in previous researches~\cite{Bekenstein,Giannios}. 
Using the key assumption that both the physical and Einstein metrics are $\eta_{\alpha\beta}$ 
at zeroth order, it was found that $\gamma =\beta =1$ as in general relativity. 

We also obtained the important result that the Newtonian potential must be static to 
be consistent with the vecvtor equation. This result is pathological to some extent. 
This also happens in Einstein-Aether theory for peculiar values of 
coupling constants~\cite{EA-PPN}. 
Because of this, we cannot determine $\alpha_{1}$ and $\alpha_{2}$. 
In relation to this, it is interesting to remember 
the instability against linear perturbation~\cite{Seifert} and occurrence of caustic 
singularities for various initial perturbations~\cite{Wiseman}. These results also support 
our pathological situation. 
Thus, as it has been pointed out in these papers~\cite{Seifert,Wiseman}, 
we should consider Einstein-Aether type generalization for the vector 
field~\cite{start} if we consider TeVeS as an alternative to dark matter. 
It is, of course, very interesting to analyze PPN parameters 
and clarify the conditions for all the PPN parameters to be satisfied. 
This is under investigation. 

Using this generalization, cosmology has been investigated in \cite{Skordis3}. 
It is also important to investigate neutron stars or black holes. 
In Einstein-Aether theory, these features are investigated and show interesting 
differences from the case in general relativity~\cite{NS,BHs,TT}. As for TeVeS, black holes 
have been investigated only for the cases where the scalar field diverges at the 
horizon~\cite{Giannios,Eva} or the scalar field is constant~\cite{Wiseman}. There is a possibility that 
the scalar field plays an important role in discussing stability of black holes as 
in black holes with Yang-Mills-Higgs fields~\cite{TT2}. Thus, we also want to consider it 
as a future subject.

%%%%%%%%%%%%%%%%%%%%%%%%%%%%%%%%%%%%%%%%%%%
\acknowledgements
%%%%%%%%%%%%%%%%%%%%%%%%%%%%%%%%%%%%%%%%%%%
We would like to thank Kei-ichi Maeda for useful discussion and for 
continuous encouragement. 

%%%%%%%%%%%%%%%%%%%%%%%%%%%%%%%%%%%%%%%%%%%%%%%%%%%%%%%%%%%%%%%%%%%%%%%%

%%%%%%%%%%%%%%%%%%%%%%%%%%%%%%%%%%%%%%%%%%%%%%%%%%%%%%%%%%%%%%%%%%%%%%%%

\end{document}